\documentclass{article}

\usepackage{geometry}
\usepackage[hidelinks]{hyperref}
\usepackage{amsfonts,latexsym}
\usepackage{amsmath}
\usepackage{amssymb}
\usepackage{mathtools}
\usepackage[ruled, vlined, linesnumbered]{algorithm2e}
\usepackage{enumitem}
\usepackage{textcomp}
\usepackage{siunitx}
\usepackage{yfonts}
\usepackage{caption}
\usepackage{tikz-cd}
\usepackage{booktabs}
\usepackage[absolute,showboxes]{textpos}

\newcommand{\email}[1]{\texttt{#1}}
\newcommand{\pf} { \noindent {\rm {\bf Proof.}} }
\newcommand{\Det}{\mathcal{D}et}
\newcommand\rres{\operatorname{Res}}
\newcommand\GL{\operatorname{GL}}
\newcommand\eva{{\rm eval_{(x \text{-} \sigma)}}}
\newtheorem{thm}{Theorem}[section]
\newtheorem{defn}[thm]{Definition}
\newtheorem{lem}[thm]{Lemma}

\newtheorem{ex}[thm]{Example}
\newtheorem{remk}[thm]{Remark}

\newcommand{\EOP} { \hfill $\Box$ }
\newcommand{\compcent}[1]{\vcenter{\hbox{$#1\circ$}}}
\newcommand{\comp}{\mathbin{\mathchoice
{\compcent\scriptstyle}{\compcent\scriptstyle}
{\compcent\scriptscriptstyle}{\compcent\scriptscriptstyle}}}

\def\A{\mbox{${\mathbb A}$}}
\def\C{\mbox{${\mathbb C}$}}
\def\D{\mbox{${\mathbb D}$}}
\def\E{\mbox{${\mathbb E}$}}

\def\Q{\mbox{${\mathbb Q}$}}

\def\Sk{\mbox{${\mathbb S}$}}
\def\Z{\mbox{${\mathbb Z}$}}

\renewcommand{\D}{\textfrak{D\,}}
\renewcommand{\mod}{ {\rm \, \, mod \, \, }}

\DeclareMathAlphabet{\mathpzc}{OT1}{pzc}{m}{it}

\NewDocumentCommand{\Eval}{>{\SplitArgument{2}{,}}m}{%
  \finalEval#1%
}
\NewDocumentCommand{\finalEval}{mmm}{%
	\emph{\rm Eval}_{(#1\text{-} #2)}%
}

\NewDocumentCommand{\Evala}{>{\SplitArgument{2}{,}}m}{%
  \finalEvala#1%
}
\NewDocumentCommand{\finalEvala}{mmm}{%
    \emph{\rm Eval}_{(#1\text{-} #2)(#3)}%
}

\newcommand\myfootnote[1]{%
  \begingroup
  \renewcommand\thefootnote{}\footnote{#1}%
  \addtocounter{footnote}{-1}%
  \endgroup
}

\TPMargin{5pt}

\newcommand{\copyrightstatement}{
\begin{textblock}{0.59}(0.20,0.79)
  \noindent \footnotesize
  Copyright~{\scriptsize \copyright} 2021 IEEE. Personal use of this material
  is permitted. Permission from IEEE must be obtained for all other uses,
  in any current or future media, including reprinting/republishing this
  material for advertising or promotional purposes, creating new collective
  works, for resale or redistribution to servers or lists, or reuse of any
  copyrighted component of this work in other workss.~\\
\end{textblock}}

\begin{document}

\title{\textbf{\Large{Resultant-based Elimination in Ore Algebra}}}
\author{Raqeeb Rasheed\\ {\normalsize Department of Computer Science,
University of Manitoba,} \\ {\normalsize Winnipeg, MB, R3T 2N2, Canada} \\
\email{\normalsize rasheedr@cs.umanitoba.ca}}
\date{}

\maketitle

\begin{abstract}
We consider resultant-based methods for elimination of indeterminates of
Ore polynomial systems in Ore algebra. We start with defining the concept
of resultant for bivariate Ore polynomials then compute it by the
Dieudonn{\'e} determinant of the polynomial coefficients. Additionally,
we apply noncommutative versions of evaluation and interpolation techniques
to the computation process to improve the efficiency of the method.

The implementation of the algorithms will be performed in Maple to evaluate
the performance of the approaches.
\\[0.5em]
\end{abstract}

\section{Introduction} \label{sec:intro}

Solving polynomial systems in general, whether in the commutative or
noncommutative case, is a fundamental mathematical interest.
In addition to computer algebra, it is used for numerous applications in
sciences, for example in engineering \cite{SymbolicEngineering},
computer vision \cite{KukelovaKSP17}, economics \cite{KUBLER2014599},
cryptology \cite{Pasalic2017}, algebraic geometry \cite{AlgGeometry18}.
Also seen in many noncommutative cases, such as in cryptography~\cite{BurgerH14},
linear codes~\cite{BOUCHER20091644}, robotics~\cite{Chirikjian2000EngineeringAO},
etc.

However, despite its popularity, the challenges associated in the
computation of large polynomial systems have limited their potential
until the advent of computer algebra systems where it became possible to
solve a large class of different type of polynomial systems. But the problem
of high computational time~\cite{Sitton2003} of such polynomial systems
remains an important concern for both researchers and industries.
Optimization techniques (such as modular methods) have been increasingly
used to improve the efficiency as the algorithms become computationally
intensive and~time \vspace{-1em}
\myfootnote{
This work has been submitted to the IEEE with the title ”Resultant-based
Elimination for Skew Polynomials”, which is an updated (and shorter) version
of this article.}

\copyrightstatement \pagebreak

\noindent consuming. This motivates the present work to develop
modular algorithms for solving noncommutative polynomial systems (by
elimination of the indeterminates).

The {\em modular approach} is a powerful method for controlling the
intermediate expression growth during the computation process that
improves the efficiency of solving the system of polynomial equations.
This method is well studied in commutative algebra, for example
in~\cite{CHEN2012610,Pavel2,Raqeeb07} in which the idea behind these
techniques is to employ resultant and gcd (greatest common divisor)
approaches to solve polynomial systems by means of {\em triangular
sets \textup(a sequence of polynomial equations with strictly increasing
main indeterminates\textup)} and then considering evaluation and
interpolation techniques to improve the efficiency of the computation
process. In this work, we propose methods to use this approach effectively for
solving multivariate (initially bivariate) Ore polynomial systems in
Ore algebra.

The transition from a commutative algebra to an Ore algebra is not a
straightforward extension, not only because of the noncommutative nature
of the Ore algebra which comes with its own challenges to navigate (compared
to the commutative case), but also because of some necessary key components
in this study are not even available on multivariate Ore polynomials in
its general form (such as multivariate resultant). Thus, we have to define
noncommutative analogues of these components to address these gaps. The
main challenges in this transition include:

\begin{enumerate}[label=$\bullet$]
	\item {\em Unavailability of multivariate resultant}:
		One of the difficulties encountered in this study is unavailability
		of multivariate resultant in Ore algebra that we could not find it
		in the literature.
	\item {\em Evaluation map is not common}: The most common evaluation map
		one thinks of (namely by plug-in values) is no longer a valid
		process in noncommutative world because it does not preserve
		multiplications.
	\item {\em Determinant is different}:
		One of the primary challenges when it comes to computing determinants
		in noncommutative algebra is the lack of a standard definition
		of determinant. Many different definitions of noncommutative
		determinant have been formulated and studied in the literature.
	\item {\em More conditions are needed for interpolation}: Selecting
		valid values for the interpolation method is different than the
		commutative case because simply using pairwise different values
		is not suitable anymore, more conditions are required.
\end{enumerate}

Although most of these challenges are known in the noncommutative algebras,
their solutions are different depending on the study.

The ultimate aim of this study is to design, optimize, and implement
symbolic algorithms for solving Ore polynomial systems. The contributions
include:

\begin{enumerate}[label=$\bullet$]
	\item New methods to eliminate indeterminates from bivariate Ore
		polynomial systems using two different ways:
		\begin{enumerate}
			\item[--] By using a direct resultant computation from the
				polynomial coefficients.
			\item[--] By applying
				noncommutative evaluation and interpolation techniques.
		\end{enumerate}
				The proof of the methods together with auxiliary lemmas
				will be provided.
	\item New definitions and properties;
		these include
		\begin{enumerate}
			\item[--] An analogous version of Sylvester's resultant for
				bivariate Ore polynomials.
			\item[--] An extended version of the operator evaluation for
				bivariate Ore polynomials.
		\end{enumerate}
	\item Design and implement the algorithms
		to demonstrate the methods.
\end{enumerate}

\noindent
{\bf Why bivariate case?} The importance of bivariate case in this study
	resides in the recursive evaluation and interpolation procedure, which
	allows us to reduce the general case of an $n\times n$ system to
	$(n-1)\times (n-1)$ system by evaluating one of the indeterminates
	then, up to details, this process can be repeated for each case
	until reaching the bivariate system which then can use the bivariate
	techniques to solve the original system.

\section{Related work} \label{sec:relatedwork}

This study is based on the concept of \emph{characteristic set
\textup(polynomials in triangular form\textup)}. This is based on
elimination theory, which is one of the core concepts in differential
and difference algebras.

Eliminations via Gr{\"o}bner bases have been studied in~\cite{Chyzak1996}
for linear differential and difference equations, followed by works
in~\cite{Mansfield2003} for the case where the indeterminates do not
commute with each other, and more recently in~\cite{Anen2014} for operator
algebra. However, these studies do not use modular approaches toward
the indeterminates.

Resultant theory is yet another technique that can be viewed as an
elimination method. Many elimination algorithms are based on resultant
computations such as; in~\cite{Tang2016} for resultant elimination using
interpolation of implicit equation, and in~\cite{Emiris2005} for computing
multivariate sparse resultant, also in~\cite{Jeronimo2004} for computing
sparse resultants using the computations of Chow forms. However,
none of these resultant methods use Ore polynomials.

The resultant for univariate skew polynomials over a division
ring has been studied in~\cite{Aleksandra2008}, and for univariate Ore
polynomials over commutative rings in~\cite{Li1996}. The subresultant
theory was also well studied for differential operators in~\cite{Chardin1991},
thereafter generalized to the univariate Ore polynomials in~\cite{Li1998AST},
and then further studied in~\cite{Hong2001} to provide an analogous version
of the subresultant expressed in terms of the solutions. Additionally,
an improved version of the subresultant sequence was studied
in~\cite{JAROSCHEK201364}. All these (sub)resultant works are studied
in the univariate case of Ore polynomials.

Differential algebra was introduced by Joseph Ritt in 1950~\cite{Ritt50},
where differential resultant was first studied for differential
operators by Ore~\cite{Ore1932} and for nonlinear differential equations
by~\cite{Ritt1932}. Since then, a variety of
studies have shown the importance of differential resultant
for example~\cite{Juan2020, Ferro1994, Zeilberger1990, Berkovich1986}
also for multivariate differential
resultants such as~\cite{Sonia2016,Rueda2010,Zwillinger1998,Ferro}.
That being said, only the {\em difference} case is considered here in this study.

Modular techniques are frequently used in computer algebra to reduce the
cost of polynomial coefficient growth. Modular algorithms for multivariate
polynomials have been most studied in the commutative case, for example
the cases of bivariate and trivariate polynomials have been studied
in~\cite{Raqeeb07}, and in a more general settings when the solution
set splits addressed in~\cite{CHEN2012610}. However, modular methods for
Ore algebra have received less attention in the literature. In~\cite{Li97},
a modular approach was used towards the coefficients of Ore polynomials
over the commutative polynomial ring $\Z[t]$. Additionally,
in~\cite{Cheng2007ModularCF} and in its Maple implementation~\cite{Labhn2014}
modular approaches have used for a matrix of Ore polynomials also over
$\Z[t]$. Recently, a modular technique on a probabilistic algorithm was
provided by~\cite{Decker2020} for computing Gr{\"o}bner bases in G-algebra
defined over the field of rational numbers $\Q$. These modular methods
are applied toward coefficients over commutative rings. None of
these studies use noncommutative evaluation and interpolation techniques.

There are some fairly recent research developments on noncommutative evaluation
and interpolation methods such as in~\cite{Zhang2010} where the Newton
interpolation for skew polynomials was studied, as well as discussed
in~\cite{Liu2014} with a different setup of evaluation values. Moreover,
a Lagrange-type interpolation with an evaluation map that vanishes at every
points over multivariate skew polynomial rings was studied
in~\cite{MARTINEZPENAS2019111}.

With the current missing studies about modular methods over Ore polynomials,
we consider extending the modular algorithms in~\cite{Raqeeb07} to
(bivariate) Ore polynomial rings.
\\[0.5em]
\textbf{Tools:} There are some packages that are implemented for Ore polynomials.
Maple package OreTools~\cite{ZLi2003} uses a modular algorithm for the
coefficients of univariate Ore polynomials over $\Z[t]$. This package,
however, does not deal with the multivariate case. The packages
Ore\_algebra~\cite{Chyzak} in Maple and ore\_algebra in
SageMath~\cite{Kauers2019} can process multivariate Ore polynomials but
they do not use modular techniques.

\section{Background}

In this section, we provide a brief review of the Ore algebra, operator
evaluation and Dieudonn{\'e} determinant which will be used in later sections.

\subsection{Ore algebra}

An Ore algebra is an algebra of noncommutative polynomials~\cite{Chyzak1996}.
It can describe linear differential, difference, and difference-differential
equations in a unified framework that collectively models a large subclass of
noncommutative rings.

In the following we recall the definition of Ore ring~\cite{Ore1933} and
followed by the definition of Ore algebra~\cite{Chyzak1996}. Note that
throughout this study $\A$ is a (skew) field, unless otherwise stated.

\begin{defn}[Ore polynomial ring]
	Let $\sigma: \A \to \A$ be an automorphism of $\A$
	\textup(called conjugate operator\textup). The noncommutative
	polynomial ring $\A[x; \sigma]$ is the set of polynomials in
	indeterminate $x$ over $\A$ with the usual polynomial addition
	$(+)$ and noncommutative multiplication defined as
	\begin{equation} \label{eq:crule}
		x a = \sigma(a) x, ~\forall a \in \A.
	\end{equation}
	This ring is called the Ore polynomial ring \textup(or left Ore
	ring\textup) with conjugate operator $\sigma$ over $\A$. Elements
	of $\A[x; \sigma]$ are called Ore polynomials.
\end{defn}

A nonzero polynomial in $\A[x;\sigma]$ has {\em degree} $n$ if $n$ is the
highest power of $x$ with a nonzero coefficient, or $-\infty$ for the
zero polynomial.

\begin{remk}
	Note that the typical commutative polynomials are a special case of
	Ore polynomials. This can be shown by taking the $\sigma$ to be the
	identity map of a commutative ring $\A$, where the commutation
	rule~{\rm (\ref{eq:crule})} simply becomes $xa=ax$ for all $a \in \A$.
	Thus, this study can be applied to the commutative case as well,
	and it will work as usual.
\end{remk}

To manage more complicated types of Ore rings than just the ring of
polynomials in one indeterminate, the process of Ore polynomial ring
can be iterated with different suitable $\sigma$ to construct Ore
polynomial rings with $n>1$ indeterminants. We state the following
definition of {\em Ore algebra}~\cite{Chyzak1996} in the bivariate
case.

\begin{defn}[Ore algebra] \label{defn:OreAlgebra}
	An Ore algebra is the iterated Ore polynomial ring
	\[
		\Sk = \A[x_1;\sigma_1][x_2;\sigma_2],
	\]
	with two commuting indeterminates $x_1,x_2$ over $\A$ and two
	automorphisms $\sigma_1, \sigma_2$ of $\Sk$ that satisfy the relations
	\begin{equation} \label{eq:eqOree}
		\sigma_j(x_i)=x_i (i \neq j), \sigma_j \sigma_i = \sigma_i \sigma_j
		~and~ x_ia=\sigma_i(a)x_i,
	\end{equation}
	for all $a \in \A$ and $1 \leq i , j \leq 2$. Elements of $\Sk$ are
	called bivariate Ore polynomials.
\end{defn}

\subsection{Operator evaluation} \label{subsec:OptEval}

One of the main differences when it comes to the evaluations in noncommutative
rings is that evaluation maps behave quite differently comparing to the
commutative case (plug-in values) mainly because noncommutative evaluation
maps generally do not preserve the products.

We are looking for a suitable evaluation map that will not only be a ring
homomorphism but also correctly handles the commutation between indeterminates
(since in this study the indeterminates are assumed to commute with each other).
For example, in $\C[x_1;\sigma_1][x_2;\sigma_2]$ over complex numbers $\C$,
substituting $i$ for $x_1$ in a multivariate expression such as $x_1 x_2$
will be $ix_2$, while the same substitution for its commuting expression
$x_2 x_1$ leads to $x_2i = \sigma_2(i)x_2$, and the two results are not
the same in general.

Although the basic concept of evaluation is the same, there are many
different noncommutative evaluation maps that serve different purposes.
In addition to the evaluating formula, there are also remainder
theorem~\cite{Cohn1985}, product formula~\cite{lam2001first}, operator
evaluation~\cite{Ulmer2013}, $\sigma$-evaluation~\cite{Zhang2015}, etc.

One of the essential parts of Ore polynomials is the presence of operators;
such as $\sigma$, in fact, all the elements of Ore polynomial rings can
be viewed as operators which naturally suggests to think of a map that
can evaluate at operators where an interesting target would be evaluating
at $\sigma$ because we know $\sigma$ is a ring homomorphism and this will
have its influence on the evaluation process. Thus, we consider what is
called the {\em operator evaluation}~\cite{Ulmer2013}.

At this point, it is convenient to state the following ring~\cite{Ulmer2013}
\begin{equation} \label{eq:diff}
	\A[\sigma; \comp]= \left\{ \sum_{i=0}^{n} a_i \sigma^i: a_i \in \A \right\},
\end{equation}

which is a left Ore ring~\cite{Ore1933} (also see~\cite[Section 3]{Milliet2018})
with the coefficient-wise addition and the composition of operators as
multiplication.
Thus, any $f = \sum_{i=0}^{n} a_i \sigma^i$ and $g =
 \sum_{j=0}^{n} b_j \sigma^j$ in $\A[\sigma; \comp]$ are added as
\[ f + g = \sum_{k=0}^{n} (a_k + b_k) \sigma^k, \]
and multiplied by the operator composition (denoted by $ f \comp g$ or
simply $f g$) as
\[ f g = \sum_{i=0}^{n} \sum_{j=0}^{n} a_i
 {\sigma^i}(b_j) {\sigma^{i+j}}. \]

We call the elements of $\A[\sigma; \comp]$ polynomials of the
indeterminate $\sigma$ over the (skew) field $\A$.

Now we can state the following Lemma related to operator evaluation for
univariate skew polynomials~\cite{Ulmer2013}.

\begin{lem} \label{lem:OperatorEval}
	Let $\textstyle f = \sum_{i=0}^{n} a_i x^i$ be an Ore polynomial
	in $\A[x;\sigma]$. The evaluation map $\Eval{x,\sigma}(f)$ defined as
	\[
	\Eval{x,\sigma}~ \colon
	\begin{array}{>{\displaystyle}r @{} >{{}}c<{{}} @{} >{\displaystyle}l}
		\A[x;\sigma] &\rightarrow& \A[\sigma; \comp]	\\
		\textstyle f = \sum_{i=0}^{n}a_ix^i	&\mapsto& \textstyle f(\sigma) =
		\sum_{i=0}^{n}a_i\sigma^i.
	\end{array}
	\]
is a morphism of rings.
\end{lem}

Accordingly, the following concepts of {\em operator evaluation} and
{\em solution} can be defined~\cite{Ulmer2013}.

\begin{defn}[Operator evaluation] \label{defn:OperatorEvala}
	Let $\textstyle f = \sum_{i=0}^{n} a_i x^i$ be an Ore polynomial
	in $\A[x;\sigma]$, and let $L_f$ denote the map $f(x)$. The operator
	evaluation $\Eval{x,\sigma}(f)$ at $a \in \A$ is defined as
	\[
	\Evala{x,\sigma,a}~ \colon
	\begin{array}{>{\displaystyle}r @{} >{{}}c<{{}} @{} >{\displaystyle}l}
		\A[x;\sigma] &\rightarrow& \A ~\\
		f = \sum_{i=0}^{n}a_ix^i	&\mapsto&
		f(\sigma)(a) =
		\sum_{i=0}^{n} a_i \sigma^i(a).
	\end{array}
	\]
	If $\Evala{x,\sigma,a}(L_f)=0$ then $y=a$ is {\em a solution} of
	$L_f(y)=0$. When the operator under evaluation is clear from context,
	we denote $f(\sigma)(a)$ by $f^{*}(a)$.
\end{defn}

\subsection{Dieudonn{\'e} determinant}

Before describing the definition of Dieudonn{\'e}
determinant~\cite{Dieudonne1943}, we would like to mention some notations
about elementary row operations in noncommutative case (mostly analogous
to the commutative case), that will be used throughout this study. For
more details on this please see, for example,~\cite[\S9.2]{Cohn2011}.

Let the notation $M_n(\Sk)$ denotes the set of all $n \times n$ matrices
over a (skew) field $\Sk$. Let $e_{ij}$ be a matrix whose elements are
all zeros except the value $1$ in the $(i,j)$-$th$ coordinate. Also,
let $E_{ij}$ represents interchanging rows $i$ and $j$. Additionally,
denote by $E_{i}(u) = I + (u-1)e_{ii}$ the elementary operation of
multiplying row $i$ by a scalar unit $u$. Finally, the {\em elementary
matrix} of adding ($q$ times row $i$) to row $j$ is denoted by
\begin{equation} \label{eq:mul_op}
	E_{ij}(q) = I + qe_{ji} ~~ (i \neq j, \, q \in \Sk).
\end{equation}

Let P be a general {\em permutation matrix} obtained from the $n \times n$
identity matrix $I$ by permuting some rows (or equivalently some columns)
according to a permutation map $\mathpzc{p}$ of $n$ elements
\[\mathpzc{p}:\{1,\ldots,n\} \rightarrow \{1,\ldots,n\}. \]

Note that the determinant $\Det(P)$, the Dieudonn{\'e} determinant which
will be described in Definition~\ref{defn:DieudonneDeterminant} and
Remark~\ref{remk:DP}, is always either $1$ or $-1$ depending on the number
of permutations whether even or odd. Sometimes, it is needed to keep the
value of $\Det(P)$ to be always 1, and in this case, it is convenient to
define what is called {\em signed permutation}, which is the same as the
general permutation matrix except whenever we permute two rows (or columns)
then we change the sign of one of them~\cite[p. 350]{Cohn2011}. Furthermore,
it is known that the signed permutation of two rows $i$ and $j$ can be
accomplished by $E_{ij}(u)$, in three steps in the form
\begin{equation} \label{eq:three_steps}
 P=E_{ij}(1)E_{ji}(-1)E_{ij}(1),
\end{equation}
which replaces row $j$ by row $i$ and row $i$ by ($-1$ times row $j$).
For example, if we apply it to the $2 \times 2$ identity matrix $I$, we
obtain
$ \begin{psmallmatrix} \, 0 & ~-1 \vspace{4pt} \\ 1 & ~0 \end{psmallmatrix}$,
noting that the determinant value stays invariant.

We can use these row operations to reduce a matrix into a simpler form
without compromising its properties and that is what we will be practicing
throughout this study. For more details about these notations and their
properties, please see~\cite[\S9.2]{Cohn2011} or~\cite[Chapter IV]{Artin2011}.

In the following section we recall the definition of the Dieudonn{\'e}
determinant via diagonal matrices, for example see~\cite[Chapter IV]{Artin2011}
or~\cite[\S20]{Draxl1983}.

\begin{defn}[Dieudonn{\'e} determinanant] \label{defn:DieudonneDeterminant}
	Let $\Sk$ be a \textup(skew\textup) field, $M_n(\Sk)$ be the $n \times n$
	matrices over $\Sk$, and $\GL_n(\Sk)$ be the multiplicative group of
	invertible $n \times n$ matrices with entries in $\Sk$. Additionally, let
	$[\Sk^{\times},\Sk^{\times}]$ be the \textup(normal\textup) multiplicative
	commutator subgroup of $\Sk$. The Dieudonn{\'e} determinant
	\textup(denoted by $\Det(A)$\textup) of a square diagonal matrix
	$A \in M_n(\Sk)$ with diagonal entries $d_i$ \textup($i = 1, \ldots, n$\textup)
	is defined as
	\begin{equation} \label{eq:det-equ1}
		\Det(A) =
			\begin{cases}
				0 & A \notin \GL_n(\Sk)   \\
				[\, {\rm sign}(\mathpzc{p}) \displaystyle
				\prod_{i=1}^{n} d_i]  & A \in \GL_n(\Sk)
			\end{cases}
	\end{equation}
	where $\mathpzc{p}$ is a permutation map and $[\cdot]$ is the canonical
	projection from $\Sk^{\times}$ to
	$\Sk^{\times}\hspace{-.12cm}/[\Sk^{\times},\Sk^{\times}].$
\end{defn}

The Ore polynomial ring $\A[x;\sigma]$ can be embedded into a skew field
$\Sk=\A(x;\sigma)$ which is the field of left fractions of
$\A[x;\sigma]$~\cite{Ore1931} (also see~\cite[Corollary 0.7.2]{Cohn2006}).
Accordingly, we can find the determinant $\Det(A)$ of any invertible matrix
$A$ over $\A[x;\sigma]$ where in Lemma~\ref{lem:DetIsPoly} we will show that
this determinant can be represented by a polynomial belong to $\A[x;\sigma]$
modulo commutators.

Note that the meaning of $\Det(A)$ in~(\ref{eq:det-equ1}) is that the Dieudonn{\'e}
determinant of an invertible matrix $A$ is a value in an abelian group
$\Sk^{\times}\hspace{-.12cm}/[\Sk^{\times},\Sk^{\times}]$, on the other
hand if $A$ is not invertible then this simply means $\Det(A)=0$. For more
details on Dieudonn{\'e} determinant and its properties please see, for
example, ~\cite{Oki2019,Artin2011,Dieudonne1943}.

\begin{remk} \label{remk:DP}
	It is known~\cite[p. 352]{Cohn2011} that for any invertible matrix
	$A$, a signed permutation matrix $P$ satisfies the property
	\begin{equation}
		\Det(PA)=\Det(A).
	\end{equation}
	Thus, using signed permutation, we can clear {\rm sign($\mathpzc{p}$)}
	in formula~(\ref{eq:det-equ1}) for invertible diagonal matrices and write
	\begin{equation} \label{eq:det-equ2}
		\Det(A) = [ \, \prod_{i=1}^{n} d_{i} \, ].
	\end{equation}
\end{remk}

One of the main properties of Dieudonn{\'e} determinant is that it satisfies
the elementary row operations~\cite{Artin2011}, analogous to the commutative
case. Therefore, Dieudonn{\'e} determinant can be computed by a procedure
similar to the Gauss-Jordan elimination method aiming to diagonalize the
matrix first and then applying the determinant formula~(\ref{eq:det-equ2}).

Furthermore, the method of triangularization of a square matrix may also
be performed to compute the Dieudonn{\'e} determinant, especially if we
would like to have the determinant value to stay in the same ring of the
matrix entries, which is the focus of Section~\ref{subsec:DDore}.

Note that Dieudonn{\'e} determinant is unique only up to commutators. We
can illustrate this with a $2 \times 2$ matrix $A$ over a (skew) field $\A$ as
\[ A =
	\begin{pmatrix}
		a_{11} &~ \,  a_{12}  \\
		a_{21}  &~ ~ a_{22}
	\end{pmatrix}.
\]
Assume $a_{11} \neq 0$ and $a_{21} \neq 0$ (otherwise the case is trivial).

\noindent
If we fix $a_{11}$ then by using row reduction properties, we can have

\[ A =
	\begin{pmatrix}
		a_{11} &~ \,  a_{12}  \\
		0  &~ ~ a_{22} - a_{21} a_{11}^{-1} a_{12}
	\end{pmatrix},
\]

\noindent
the Dieudonn\'e determinant is
\begin{equation} \label{eq:q1}
	[a_{11} a_{22} - a_{11} a_{21} a_{11}^{-1} a_{12}].
\end{equation}

\noindent
On the other hand, if we fix $a_{21}$ then

\[ A =
	\begin{pmatrix}
		0  &~ ~ \,  a_{12} - a_{11} a_{21}^{-1} a_{22}  \\
		a_{21}  &~ ~ a_{22}
	\end{pmatrix},
\]
thus the Dieudonn\'e determinant in this case is
\begin{equation} \label{eq:q2}
	[a_{21} a_{11} a_{21}^{-1} a_{22} - a_{21} a_{12}].
\end{equation}

\noindent
Now we need to prove that (\ref{eq:q1}) and (\ref{eq:q2}) are equivalent.
The following diagram illustrates the proof:

\pagebreak

\begin{center}
	{\scriptsize ?} \\
	\vspace{-.5em}
	$[a_{11} a_{22} - a_{11} a_{21} a_{11}^{-1} a_{12}]$  ~ $\equiv$ ~
	$[a_{21} a_{11} a_{21}^{-1} a_{22} - a_{21} a_{12}]$ \\
	\vspace{.5em}
	\hspace{-1em}  $\Downarrow$  \hspace{12em}  $\Downarrow$  \\
	\vspace{.5em}
	$[a_{11} a_{22} - a_{11} a_{21} a_{11}^{-1} a_{21}^{-1} a_{21} a_{12}]$  ~
	$~  ~ $ ~   $[a_{21} a_{11} a_{21}^{-1} a_{11}^{-1} a_{11} a_{22} - a_{21} a_{12}]$ \\
	\vspace{.5em}
	\hspace{-1em}  $\Downarrow$  \hspace{12em}  $\Downarrow$  \\
	\vspace{.5em}
	\hspace{2em} $[a_{11} a_{22} - x a_{21} a_{12}]$  ~ $~  ~ $ ~
	$[x^{-1} (a_{11} a_{22} - x a_{21} a_{12})]$ \\
\end{center}

where $x = a_{11} a_{21} a_{11}^{-1} a_{21}^{-1}$ which is a commutator. Thus,
it shows the equivalence and this means the determinant is well defined (modulo
commutators).

\begin{remk} \label{remk:ModCommutators}
	Note that in this study, all computations of Dieudonn{\'e} determinant
	being performed modulo commutators even if this is not explicitly stated.
\end{remk}

\section{Elimination}

In this section we discuss two different but related new methods
for eliminating indeterminates in a system of Ore polynomials,
one based on the direct computation of resultant
from the polynomials coefficients, and the other is based on the
evaluation and interpolation techniques. First,
we need to have a property that Dieudonn{\'e} determinant
is a unique {\em polynomial} (modulo commutators) when the entries are
Ore polynomials as in the following section.

\subsection{Dieudonn{\'e} determinant over Ore polynomial rings} \label{subsec:DDore}

From the perspective of normal forms of matrices, one can think of
diagonalizing an invertible matrix $A$ over $\A[x;\sigma]$ to be
converted to the form:
\begin{equation} \label{eq:deteq3}
	D = UAV = {\rm diag}(1,\cdots,1,d),
\end{equation}

where $U$ and $V$ are {\em unimodular matrices}, $D$ is a diagonal matrix
whose diagonal entries are all  1\textquotesingle s except the last entry
which is equal to $d \in \A(x;\sigma)$, that means we can identify the
matrix $D$ by an element $d$ in $\A(x;\sigma)$.

Using formula~(\ref{eq:det-equ2}), the Dieudonn{\'e} determinant of $D$ as
in~(\ref{eq:deteq3}) may be obtained by simply multiplying the diagonal
entries;
\begin{equation} \label{eq:deteq4}
	\Det(D) = [d].
\end{equation}
However, one problem of this
process for our study is that, in general, the value $d$ becomes a rational
function in $\A(x;\sigma)$ rather than staying a polynomial in $\A[x;\sigma]$,
something that we would like to address in this section.

Knowing that the Dieudonn{\'e} determinant of {\em an upper \textup(or
lower\textup) triangular matrix} is the product of the diagonal entries
\textup(see for example,~\cite{Sylvain2019} p. 822,~\cite{Draxl1979}
p. 3 \, Example 2\textup), we can reduce a matrix to a triangular form
in a manner that all the diagonal entries become polynomials so that
their product is again a polynomial in $\A[x;\sigma]$. This can be
obtained by using elementary row operations over a Euclidean domain,
as in Lemma~\ref{lem:DetIsPoly}, that we prove it similar
to~\cite{Giesbrecht2013}, but using different techniques. The next
lemma is perhaps well known but we couldn't find a precise reference
for it; we prove it anyway, for completeness and verification purposes.

\begin{lem} \label{lem:DetIsPoly}
	Let $\A[x;\sigma]$ be an Ore polynomial ring. Dieudonn{\'e} determinant
	of an invertible matrix $A \in \A[x; \sigma]^{n \times n}$
	can be represented by a unique Ore
	polynomial (modulo commutators).
\end{lem}

\noindent
\pf Let $A$ be an invertible matrix in $\A[x; \sigma]^{n \times n}$,
suppose $B=A$. For each $i=1,\ldots,n-1$, we can assume $i$-th column of
the matrix $B$ is non-zero (otherwise the determinant is 0), select a
non-zero entry below the diagonal element with the lowest degree in the
same $i$-th column (move to the next $i$ if all entries below the diagonal
element are zeros), then by using the row operation~(\ref{eq:three_steps})
we can interchange it with the $(i,i)$-th value (if not already).
As our ring of polynomials is a Euclidean domain~\cite{Ore1933}, for each
element $b_{ki}$ $(k>i)$ of $B$ if $\deg(b_{ki}) > \deg(b_{ii})$ then there
exist $q, r \in \A[x;\sigma]$ such that $b_{ki} - q b_{ii} = r$ where
$\deg(r) < \deg(b_{ii})$, and thus we can use the row operation~(\ref{eq:mul_op})
to replace the entry $b_{ki}$ below the diagonal by a smaller degree $r$,
continue in this manner until eventually the zero degree (and then the zero
value) will be reached for all $k>i$ in the $i$-th column (other than the
one exactly on the diagonal line).

Repeating this process until all the $i$-columns ($i=1,\ldots, n-1$) are
checked yields to the desired matrix. Note that Euclidean domain allow us
to keep the diagonal elements $b_{ii}$ of $B$ remain polynomials in
$\A[x;\sigma]$.

Therefore, by applying~(\ref{eq:det-equ2}) we conclude that the Dieudonn{\'e}
determinant can be written as the product of the diagonal entries
\[\Det(A) = \Det(B) = {[\, \prod_{i=1}^{n} b_{ii} \, ].}\]
\EOP

\begin{remk}
	There is no specific reason for selecting an upper triangular form
	in Lemma~\ref{lem:DetIsPoly}. Indeed, we can use as well a lower triangular
	form with a similar proof by applying elementary row operations from
	the right \textup(column operations\textup). Which means any invertible
	matrix with Ore polynomial entries can be {\em diagonalized} with Ore
	polynomial entries on the main diagonal, simply by making it upper
	and lower triangular.
\end{remk}

Another useful property of the Dieudonn{\'e} determinant is that it does
not depend on the choice of elementary row operations, neither on the
order in the product of diagonal entries of the matrix (see for
example,~\cite{Sylvain2019} p. 822).

\subsection{Elimination by direct resultant} \label{sec:methodology}

Resultant is a powerful tool in the theory of elimination which is commonly
used for solving systems of polynomial equations (due to its lower
computational complexity compared to Gr{\"o}bner-based methods).

We provide the following new definition of resultant of two bivariate
Ore polynomials in an Ore algebra that can compute the resultant
directly from the polynomials coefficients (generalizing the commutative
case of Sylvester's determinant).

\begin{defn} \label{defn:OperatorBivarResEval}
	Let $\Sk=\A[x_1;\sigma_1][x_2;\sigma_2]$
	be an Ore algebra. Consider two bivariate polynomials
	$f=\sum_{i=0}^{n} a_i(x_1) {x_2^i}$ and $g=\sum_{j=0}^{m} b_j(x_1) {x_2}^j$
	in $(\A[x_1;\sigma_1])[x_2;\sigma_2]$ where $a_i(x_1), b_j(x_1) \in
	\A[x_1;\sigma_1]$.
	We define the resultant of $f$ and $g$ with respect to $x_2$ \textup(denoted by
	$\rres_{x_2}(f,g)$\textup), by the following Dieudonn{\'e} determinant
	\[\rres_{x_2}(f,g)= \hspace{.1em}
	\begin{matrix}
		x_2^{m-1}f \\
		\vspace{-.4em} \\
		x_2^{m-2}f \\
		\vspace{1.2em}
		\vdots \\
		~~~~~f \\
		\vspace{1.5em} x_2^{n-1} g \\[.1em]
		\vdots \\
		\ \\
		\vspace{-1.3em} \\
		~x_2 \, g \\
		~~~~ \, g \\
	\end{matrix}
	\left |
	\arraycolsep=1.4pt\def\arraystretch{1.48}
	\begin{array}{cccccccc}
		  a_n^{\scalebox{.55}{\rm [m-1]}} {\scalebox{.90}{\rm ($x_1$)}}&a_{n-1}^{\scalebox{.55}{\rm
		  [m-1]}} {\scalebox{.90}{\rm ($x_1$)}}  &\cdots&\cdots&a_0^{\scalebox{.55}{\rm [m-1]}} {\scalebox{.90}{\rm ($x_1$)}}&&&\\
		  &a_n^{\scalebox{.55}{\rm [m-2]}} {\scalebox{.90}{\rm ($x_1$)}}&a_{n-1}^{\scalebox{.55}{\rm [m-2]}} {\scalebox{.90}{\rm ($x_1$)}}&\cdots&\cdots&a_0^{\scalebox{.55}{\rm [m-2]}} {\scalebox{.90}{\rm ($x_1$)}}&&\\
	  &&&&\ddots&&&\\
		  &&&a_n^{\scalebox{.55}{\rm [0]}} {\scalebox{.90}{\rm ($x_1$)}}&a_{n-1}^{\scalebox{.55}{\rm [0]}} {\scalebox{.90}{\rm ($x_1$)}}&\cdots&\cdots&a_0^{\scalebox{.55}{\rm [0]}} {\scalebox{.90}{\rm ($x_1$)}}\\
		  b_m^{\scalebox{.55}{\rm [n-1]}} {\scalebox{.90}{\rm ($x_1$)}}&b_{m-1}^{\scalebox{.55}{\rm [n-1]}} {\scalebox{.90}{\rm ($x_1$)}}&\cdots&b_0^{\scalebox{.55}{\rm [n-1]}} {\scalebox{.90}{\rm ($x_1$)}}&&&&\\
		  &b_m^{\scalebox{.55}{\rm [n-2]}} {\scalebox{.90}{\rm ($x_1$)}}&b_{m-1}^{\scalebox{.55}{\rm [n-2]}} {\scalebox{.90}{\rm ($x_1$)}}&\cdots&b_0^{\scalebox{.55}{\rm [n-2]}} {\scalebox{.90}{\rm ($x_1$)}}&&&\\
		  &&&&\ddots&&&\\
		  &&&b_m^{\scalebox{.55}{\rm [1]}} {\scalebox{.90}{\rm ($x_1$)}}&b_{m-1}^{\scalebox{.55}{\rm [1]}} {\scalebox{.90}{\rm ($x_1$)}}&\cdots&b_0^{\scalebox{.55}{\rm [1]}} {\scalebox{.90}{\rm ($x_1$)}}&\\
		  &&&&b_m^{\scalebox{.55}{\rm [0]}} {\scalebox{.90}{\rm ($x_1$)}}&b_{m-1}^{\scalebox{.55}{\rm [0]}} {\scalebox{.90}{\rm ($x_1$)}}&\cdots&b_0^{\scalebox{.55}{\rm [0]}} {\scalebox{.90}{\rm ($x_1$)}}
	\end{array}\right |, \nonumber \\
	\]
	where the $i$-th row \textup($i=1,\ldots,m$\textup) contains the
	coefficient sequence of the multiplication $x_2^{m-i}f$, the coefficients
	of this multiplication are denoted by
	$a_j^{\scalebox{.55}{\rm [m-i]}} {\scalebox{.90}{\rm ($x_1$)}}$
	\textup($j=n,\ldots,0$\textup). Similarly, the $(m+i)$-th row
	\textup($i=1,\ldots,n$\textup), contains the coefficients of
	$x_2^{n-i}g$, these coefficients are denoted by
	$b_j^{\scalebox{.55}{\rm [n-i]}} {\scalebox{.90}{\rm ($x_1$)}}$
	\textup($j=m,\ldots,0$\textup). Thus, for notational simplicity, we
	can write the resultant $\rres_{x_2}(f,g)$ in the form
	\begin{equation} \label{eq:equ_thm}
		\rres_{x_2}(f,g) =
		\Det(x_2^{m-1} f, \ldots, x_2 f, \, f, \,\,x_2^{n-1} g, \ldots, x_2 g, \, g). \\
	\end{equation}
\end{defn}

Recall that the indeterminates $x_1$ and $x_2$ do not commute with the
coefficients but rather act according to the ring automorphisms
$\sigma_1$ and $\sigma_2$ such that for each $a \in \A$,
\begin{equation}
	x_1 a = \sigma_1(a) x_1 {\rm ~~and~~} x_2 a = \sigma_2(a) x_2,
\end{equation}

which means the noncommutative properties in the determinants' entries
are preserved, since the rows are multiplied by a power of $x_2$ to the
left.

\begin{remk}
	The definition of Dieudonn{\'e} determinant~\ref{defn:DieudonneDeterminant}
	requires entries to be in a (skew) filed, this can be obtained by embedding
	$\A[x_1,\sigma_1]$ in a (skew) filed~\cite[Corollary 0.7.2]{Cohn2006}
	(since $\A[x_1,\sigma_1]$ is an Ore ring).
\end{remk}

\subsection{Elimination by evaluation and interpolation resultant}

In this section we describe another method to compute resultant with Ore
polynomial entries by evaluation and interpolation techniques (generalizing
the commutative case of~\cite{Raqeeb07}). We prove a theorem which
establishes a relation between bivariate resultants and evaluation maps.

Additionally, we apply the operator evaluation methods in~\cite{Eric2020}
or~\cite{Caruso2017} to improve the efficiency of the polynomial
multiplications during the computation of the resultant, but we proceed
slightly differently by considering the operator $\sigma_1$ itself as
an element of the base ring (as in Remark~\ref{remk:ext}), which provides
a more convenient way to determine the commutators and the conjugacy
classes that will be needed later.

\begin{remk} \label{remk:ext}
	Let $\A[\sigma_1,\comp]$ be the left Ore ring as in
	\textup({\rm \ref{eq:diff}}\textup), then we
	can have the ring $\A[\sigma_1,\comp][x_1;\sigma_1]$ which is also a left Ore
	ring~\cite[Corollary 1]{Chyzak1996} with the properties
	\begin{equation} \label{eq:extrOre}
		x_1 \sigma_1 = \sigma_1 x_1, ~\sigma_1 a  = \sigma_1(a) \sigma_1
		~{\rm and}~  x_1 a = \sigma_1(a) x_1,  ~\forall a \in \A.
	\end{equation}
	In the same manner, this process can be iterated to have the Ore algebra
	\[\A[\sigma_1;\comp][x_1;\sigma_1][x_2;\sigma_2].\]
	Such that, in addition to the relations~\textup(\ref{eq:eqOree}\textup)
	and~\textup(\ref{eq:extrOre}\textup), it satisfies
	\begin{equation}
		\sigma_1 x_2 = x_2 \sigma_1.
	\end{equation}
	Since $\A[\sigma_1; \comp]$ is a left Ore ring, it can be embedded into
	a (skew) field $\tilde{\A}=\A(\sigma_1,\comp)$~\cite[Corollary 0.7.2]{Cohn2006},
	for which (later in Definition~\ref{defn:OperatorEval_2}) we need a
	multiplicative group, denoted by $\tilde{\A}^{\times}$, that
	contains nonzero elements of $\A(\sigma_1,\comp)$. The multiplication
	in $\tilde{\A}^{\times}$ is defined similarly to the multiplication with
	rational functions in Ore algebra where for any two left fractions
	$f_2^{-1} f_1$ and $g_2^{-1} g_1 in \tilde{\A}^{\times}$, the multiplication
	is defined as
	\[ f_2^{-1}f_1\,.\,g_2^{-1}g_1 = (d_2f_2)^{-1} d_1g_1 \label{eq:2}, \]
	such that $d_1 g_2 = d_2 f_1$, where $d_1$ and $d_2$ are from left
	Ore conditions\footnote{By the left Ore condition we can find $d_1, d_2 \in
	\tilde{\A}$ such that  $d_1 g = d_2 f$,  for all nonzero elements $f,
	g \in \tilde{\A}$.}.
\end{remk}

\noindent
From the Remark~\ref{remk:ext}, we can conclude the following:
\begin{lem} \label{lem:sigmax2}
	Let $\tilde{\A}[x_1;\sigma_1][x_2;\sigma_2]$ be an Ore algebra where
	$\tilde{\A} = \A(\sigma_1,\comp)$. For all nonnegative integers
	$n$,$m$ and for all $a \in \A$, the following holds:
\end{lem}
\begin{enumerate}[align=left, label=\textup(\roman*\textup)]
	\item \label{lem:i} $\sigma_1^n x_2^m = x_2^m \sigma_1^n$
	\item \label{lem:ii} $\sigma_1^n a \sigma_1^m =
		{\sigma_1^n}(a)\sigma_1^{n+m}.$
\end{enumerate}

\noindent
Note that, in this Lemma~\ref{lem:sigmax2}, the product~\ref{lem:i} is
commutative while~\ref{lem:ii} is not.

\begin{ex}
	Consider the Ore algebra $\tilde{\A}[x_1;\sigma_1][x_2;\sigma_2]$ where
	$\tilde{\A} = \A(\sigma_1;\comp)$, we can find $(\sigma_1 x_2)(a)$ as
	following:
	\begin{align*}
		(\sigma_1 x_2)(a) &= (x_2 \sigma_1)(a) \nonumber = x_2 \,
		\sigma_1(a) = \sigma_2(\sigma_1 (a)) x_2.
	\end{align*}
	Similarly for all $a,b \in \A$, we can describe $(\sigma_1 a)(b)$ as:
	\begin{align*}
		(\sigma_1 a)(b) &= (\sigma_1(a) \sigma_1)(b) = \sigma_1(a) \sigma_1(b) =
		\sigma_1(ab).  \nonumber
	\end{align*}
\end{ex}

In the following, we extend the evaluation notations described in
Section~\ref{subsec:OptEval} to {\em bivariate} Ore polynomials as defined below:

\begin{defn} \label{defn:OperatorBivarEval}
	Let $\tilde{\A}[x_1;\sigma_1][x_2;\sigma_2]$ be a bivariate Ore polynomial
	ring. Since we have commuting indeterminates $x_1$ and $x_2$, we can
	consider the Ore ring in the form $\Sk=\E[x_1;\sigma_1]$ where
	$\E=\tilde{\A}[x_2;\sigma_2]$, that is, polynomials are regarded with
	respect to $x_1$. Then for each polynomial $\textstyle f = \sum_{i=0}^{n}
	\alpha_i {x_1^i}$, $ \alpha_i \in \E$, we define the evaluation map
	$\Eval{x_1,\sigma_1}(f)$ as
	\[
	\Eval{x_1,\sigma_1}~ \colon
	\begin{array}{>{\displaystyle}r @{} >{{}}c<{{}} @{} >{\displaystyle}l}
		\E[x_1;\sigma_1] &\rightarrow& \E \\
		f = \sum_{i=0}^{n}\alpha_i{x_1^i} &\mapsto& f(x_2,\sigma_1) =
		\sum_{i=0}^{n}\alpha_i {\sigma_1^i}.
	\end{array}
	\]
\end{defn}

In the following, we show that the map $\Eval{x_1,\sigma_1}$ is a ring
homomorphism (morphism of rings) for {\em bivariate} Ore polynomials.

\begin{lem} \label{lem:EvalIsMorph}
	The map $\Eval{x_1,\sigma_1}$
	is a ring homomorphism.
\end{lem}

\pf Since we have commuting indeterminates, the polynomials can be regarded
with respect to $x_1$ in $\Sk = (\tilde{\A}[x_2;\sigma_2])[x_1;\sigma_1]$.
Let $f=\sum_{i=0}^{n} \alpha_i {x_1^i}$ and $g=\sum_{j=0}^{m} \beta_j {x_1^j}$
be two polynomials in $\Sk$ with $\alpha_i,\beta_j \in \tilde{\A}[x_2;\sigma_2]$.

It is easy to check that the property of ring homomorphism holds for the
addition and identity. Now we check the multiplication:
\begingroup
\allowdisplaybreaks
\begin{align*} \label{foefe}
	\Eval{x_1,\sigma_1}(fg) &=  \Eval{x_1,\sigma_1}(\sum_{i=0}^{n} \alpha_i {x_1^i}  \sum_{j=0}^{m} \beta_j {x_1^j}) \\
							&=  \Eval{x_1,\sigma_1}(\sum_{i=0}^{n} \sum_{j=0}^{m} \alpha_i {x_1^i} \beta_j {x_1^j}) \\
							&=  \Eval{x_1,\sigma_1}(\sum_{i=0}^{n} \sum_{j=0}^{m} \alpha_i
							 {\sigma_1^i}(\beta_j) {x_1^{i+j}}) \\
							&= \sum_{i=0}^{n} \sum_{j=0}^{m}
							 \alpha_i {\sigma_1^i}(\beta_j) {\sigma_1^{i+j}} \\
							&= \sum_{i=0}^{n} \sum_{j=0}^{m}
							 \alpha_i {\sigma_1^i}(\beta_j {\sigma_1^{j}}) \\
							&= \sum_{i=0}^{n} \alpha_i {\sigma_1^i} \sum_{j=0}^{m} \beta_j {\sigma_1^{j}} \\
							&=  \Eval{x_1,\sigma_1}(\sum_{i=0}^{n} \alpha_i {x_1^i})
							 \Eval{x_1,\sigma_1}(\sum_{j=0}^{m} \beta_j {x_1^j}) \\
							&= \Eval{x_1,\sigma_1}(f) \Eval{x_1,\sigma_1}(g) \\
\end{align*}
\endgroup
\EOP

In the following, we define Eval when the input polynomial is regarded
as modulo commutators.

\begin{defn} \label{defn:OperatorEval_2}
	Let $\tilde{\A}(x;\sigma)$ be a \textup(skew\textup) field and
	let $\tilde{\A}^{\times}(x;\sigma)$ denotes multiplicative group of
	$\tilde{\A}(x;\sigma)$ containing nonzero elements of $\tilde{\A}(x;\sigma)$.
	Let $f = f_2^{-1} f_1$ be an element in $\tilde{\A}^{\times}(x;
	\sigma)/[\tilde{\A}^{\times}(x;\sigma),\tilde{\A}^{\times}(x;\sigma)]$.
	Additionally, let \D denotes the normal subgroup generated by the
	multiplicative commutators of $\tilde{\A}^{\times}(x;\sigma)$, similarly
	let $\D'$ be the commutators subgroup of $\tilde{\A}^{\times}$. The
	modular evaluation map $\Eval{x,\sigma}(f)$ is defined as:
	\[
	\Eval{x,\sigma}~ \colon
	\begin{array}{>{\displaystyle}r @{} >{{}}c<{{}} @{} >{\displaystyle}l}
		\tilde{\A}^{\times}(x; \sigma)/[\tilde{\A}^{\times}(x;\sigma),\tilde{\A}^{\times}(x;\sigma)]  &\rightarrow& \tilde{\A}^{\times}/[\tilde{\A}^{\times},\tilde{\A}^{\times}]  \\
		\textstyle f = f_2^{-1} f_1 ~\mod ~\D &\mapsto&
		\textstyle f(\sigma)=(f_2(\sigma))^{-1} f_1(\sigma) ~\mod ~\D', \, f_2(\sigma)
		\neq 0. \\
	\end{array}
	\]
	In particular, if $\textstyle f = \sum_{i=0}^{n} a_i x^i$ is an Ore
	polynomial in
	$\tilde{\A}^{\times}(x; \sigma)/[\tilde{\A}^{\times}(x;\sigma),\tilde{\A}^{\times}(x;\sigma)]$
	then
	\[
		\Eval{x,\sigma}~ \colon \textstyle f = \sum_{i=0}^{n}a_ix^i ~\mod ~\D \mapsto \textstyle f(\sigma) = \sum_{i=0}^{n}a_i\sigma^i ~\mod ~\D'.
	\]
\end{defn}

\begin{remk} \label{remk:modularByDefault}
Note that in this study, we assume the evaluation map ${\rm Eval}$ becomes
modular (by default) as in Definition~\ref{defn:OperatorEval_2} when the
input argument computed modulo commutators.
\end{remk}

In the following, we show that the map $\Eval{x,\sigma}$ is well defined.
\begin{lem}
	The evaluation map $\Eval{x,\sigma}$ as in Definition~\ref{defn:OperatorEval_2}
	is well defined.
\end{lem}
\noindent
\begin{minipage}{0.5\textwidth}
	\pf Since any quotient group by its commutator subgroup is abelian,
	$\tilde{\A}^{\times}/[\tilde{\A}^{\times},\tilde{\A}^{\times}]$ has
	an abelian structure where the {\em universal property of abelianization}
	follows as in Figure~\ref{fig:abelian}, such that
	\[\eva = \Eval{x,\sigma} \comp \pi, \]
\end{minipage}\hfill
\begin{minipage}{0.48\textwidth}
	\begin{tikzpicture}
	\matrix(m)[matrix of math nodes,
	row sep=5em, column sep=2em,
	text height=1.5ex, text depth=0.25ex]
	{
		\tilde{\A}^{\times}(x;\sigma)  &
		\tilde{\A}^{\times}/[\tilde{\A}^{\times},\tilde{\A}^{\times}] \\
		& \tilde{\A}^{\times}(x;
		\sigma)/[\tilde{\A}^{\times}(x;\sigma),\tilde{\A}^{\times}(x;\sigma)] \\
	};
	\path[->]
	(m-1-1) edge node[auto] {$\eva$} (m-1-2);
	\path[->]
	(m-1-1) edge node[left] {$\pi~~$} (m-2-2);
	\path[->]
	    (m-2-2) edge node[right] {$\Eval{x,\sigma}$} (m-1-2);
	\end{tikzpicture}
	\vspace{-2em}

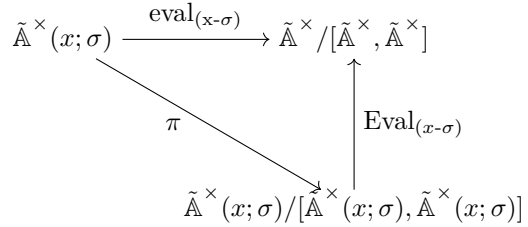
\captionof{figure}{Universal property of abelianization.}
\label{fig:abelian}
\end{minipage}
where $\pi$ is the canonical map and $\eva$ is homomorphism between
multiplicative groups defined as
\[
	\eva~ \colon f \mapsto f(\sigma) ~\mod ~\D', ~ \forall f \in \tilde{\A}^{\times}(x;\sigma).
\]

\noindent
We know that
\[
	{\rm ~ker~} \pi = [\tilde{\A}^{\times}(x;\sigma),\tilde{\A}^{\times}(x;\sigma)] \leq {\rm
~ker~} \eva.
\]

\noindent
Now, to show that $\Eval{x,\sigma}$ is well defined, suppose
	\[ f \mod \D = g \mod \D, \]
implies that
	\[ f g^{-1} \in \D = [\tilde{\A}^{\times}(x;\sigma),\tilde{\A}^{\times}(x;\sigma)]
	\leq {\rm ~ker~} \eva. \]
Thus,
\[ \eva(fg^{-1}) = \eva(f) \, \eva(g)^{-1} = 1 \mod \D', \]
that is $\eva(f)=\eva(g)$, implies that
\[ (\Eval{x,\sigma} \comp \pi)(f)  =
(\Eval{x,\sigma} \comp \pi)(g),  \]
which means $\Eval{x,\sigma}(f \mod \D) = \Eval{x,\sigma}(g \mod \D).$

\EOP

\begin{lem} \label{lem:OpToCoeff}
	Let $\Sk=\tilde{\A}[x_1;\sigma_1][x_2;\sigma_2]$ be an Ore algebra.
	For any polynomial $f=\sum_{i=0}^{n} a_i(x_1) {x_2^i}$ in
	$\Sk$ where $a_i(x_1) \in \tilde{\A}[x_1;\sigma_1]$, we have
	\[ \Eval{x_1,\sigma_1}(f)=\sum_{i=0}^{n} a_i(\sigma_1) {x_2^i}.\]
\end{lem}
\pf Let $f=\sum_{i=0}^{n} a_i(x_1) {x_2^i}$ be a polynomial in
$\tilde{\A}[x_1;\sigma_1][x_2;\sigma_2]$. Since we have commuting indeterminates
$x_1$ and $x_2$, we can consider the ring
$\Sk = (\tilde{\A}[x_2;\sigma_2])[x_1;\sigma_1]$ where the polynomial $f$ can be
arranged in the form $f=\sum_{j=0}^{m} \alpha_j(x_2) {x_1^j}$,
$\alpha_j(x_2) \in \tilde{\A}[x_2;\sigma_2]$. By Definition~\ref{defn:OperatorBivarEval}
of ${\rm Eval_{(x_1 \text{-} \sigma_1)}}$, we
can have $\Eval{x_1,\sigma_1}(f)=\sum_{j=0}^{m} \alpha_j(x_2) {\sigma_1^j}$.
By Lemma~\autoref{lem:sigmax2}-\ref{lem:i},
we can re-arrange the evaluated polynomial in the form
\[ \Eval{x_1,\sigma_1}(f)=\sum_{i=0}^{n} a_i(\sigma_1) {x_2^i}.\]

\EOP

A similar technique of Definition~\ref{defn:OperatorBivarResEval} for resultant
when the entries contain automorphisms can be described as following.

\begin{defn} \label{defn:OperatorBivarResEval2}
	Let $\Sk=\tilde{\A}[x_1;\sigma_1][x_2;\sigma_2]$
	be an Ore algebra. Consider two bivariate polynomials
	$f=\sum_{i=0}^{n} a_i(x_1) {x_2^i}$ and $g=\sum_{j=0}^{m} b_j(x_1) {x_2}^j$
	in $(\tilde{\A}[x_1;\sigma_1])[x_2;\sigma_2]$ where $a_i(x_1), b_j(x_1) \in
	\tilde{\A}[x_1;\sigma_1]$. By Lemma~\ref{lem:OpToCoeff}, we can let
	$f'=\Eval{x_1,\sigma_1}(f)=\sum_{i=0}^{n} a_i(\sigma_1) {x_2^i}$ and
	$g'=\Eval{x_1,\sigma_1}(g)=\sum_{j=0}^{m} b_j(\sigma_1) {x_2}^j$.
	Accordingly, we define
	the resultant of $f'$ and $g'$ with respect to $x_2$ \textup(denoted by
	$\rres_{x_2}(f',g')$\textup)  by the following Dieudonn{\'e} determinant;
	\[\rres_{x_2}(f',g')= \hspace{.1em}
		\begin{matrix}
		x_2^{m-1}f' \\
		\vspace{-.4em} \\
		x_2^{m-2}f' \\
		\vspace{.5em} \\
		\vdots \\
		~~~~~f' \\
		\vspace{1em} x_2^{n-1} g' \\[.1em]
		\vdots \\
		\ \\
		\vspace{0.3em} \\
		~x_2 \, g' \\
		~~~~ \, g' \\
		\end{matrix}
	\left |
	\arraycolsep=1.4pt\def\arraystretch{1.75}
	\begin{array}{cccccccc}
		  a_n^{\scalebox{.55}{\rm [m-1]}} {\scalebox{.75}{\rm ($\sigma_1$)}}&a_{n-1}^{\scalebox{.55}{\rm
		  [m-1]}} {\scalebox{.75}{\rm ($\sigma_1$)}}  &\cdots&\cdots&a_0^{\scalebox{.55}{\rm [m-1]}} {\scalebox{.75}{\rm ($\sigma_1$)}}&&&\\
		  &a_n^{\scalebox{.55}{\rm [m-2]}} {\scalebox{.75}{\rm ($\sigma_1$)}}&a_{n-1}^{\scalebox{.55}{\rm [m-2]}} {\scalebox{.75}{\rm ($\sigma_1$)}}&\cdots&\cdots&a_0^{\scalebox{.55}{\rm [m-2]}} {\scalebox{.75}{\rm ($\sigma_1$)}}&&\\
	  &&&&\ddots&&&\\
		  &&&a_n^{\scalebox{.55}{\rm [0]}} {\scalebox{.75}{\rm ($\sigma_1$)}}&a_{n-1}^{\scalebox{.55}{\rm [0]}} {\scalebox{.75}{\rm ($\sigma_1$)}}&\cdots&\cdots&a_0^{\scalebox{.55}{\rm [0]}} {\scalebox{.75}{\rm ($\sigma_1$)}}\\
		  b_m^{\scalebox{.55}{\rm [n-1]}} {\scalebox{.75}{\rm ($\sigma_1$)}}&b_{m-1}^{\scalebox{.55}{\rm [n-1]}} {\scalebox{.75}{\rm ($\sigma_1$)}}&\cdots&b_0^{\scalebox{.55}{\rm [n-1]}} {\scalebox{.75}{\rm ($\sigma_1$)}}&&&&\\
		  &b_m^{\scalebox{.55}{\rm [n-2]}} {\scalebox{.75}{\rm ($\sigma_1$)}}&b_{m-1}^{\scalebox{.55}{\rm [n-2]}} {\scalebox{.75}{\rm ($\sigma_1$)}}&\cdots&b_0^{\scalebox{.55}{\rm [n-2]}} {\scalebox{.75}{\rm ($\sigma_1$)}}&&&\\
		  &&&&\ddots&&&\\
		  &&&b_m^{\scalebox{.55}{\rm [1]}} {\scalebox{.75}{\rm ($\sigma_1$)}}&b_{m-1}^{\scalebox{.55}{\rm [1]}} {\scalebox{.75}{\rm ($\sigma_1$)}}&\cdots&b_0^{\scalebox{.55}{\rm [1]}} {\scalebox{.75}{\rm ($\sigma_1$)}}&\\
		  &&&&b_m^{\scalebox{.55}{\rm [0]}} {\scalebox{.75}{\rm ($\sigma_1$)}}&b_{m-1}^{\scalebox{.55}{\rm [0]}} {\scalebox{.75}{\rm ($\sigma_1$)}}&\cdots&b_0^{\scalebox{.55}{\rm [0]}} {\scalebox{.75}{\rm ($\sigma_1$)}}
	\end{array}\right |, \nonumber \\
	\]
	where the $i$-th row \textup($i=1,\ldots,m$\textup) contains the
	coefficient sequence of the multiplication $x_2^{m-i}f'$, the coefficients
	of this multiplication are denoted by $a_j^{\scalebox{.55}{\rm [m-i]}}
	{\scalebox{.75}{\rm ($\sigma_1$)}}$
	\textup($j=n,\ldots,0$\textup).
	Similarly, the $(m+i)$-th row \textup($i=1,\ldots,n$\textup), contains
	the coefficients of $x_2^{n-i}g'$, these coefficients are denoted by
	$b_j^{\scalebox{.55}{\rm [n-i]}} {\scalebox{.75}{\rm ($\sigma_1$)}}$
	\textup($j=m,\ldots,0$\textup). An interesting observation here is
	that the noncommutative property is preserved, since the rows are
	multiplied by a power of $x_2$ to the left, which means the coefficients
	have to follow the commutation rule.
	Thus, we can write the resultant $\rres_{x_2}(f',g')$ in the form
	\[
	\rres_{x_2}(f',g')	   = \Det(x_2^{m-1} f', \ldots, x_2 f', \, f', \,\,x_2^{n-1} g', \ldots, x_2
	g', \, g'), \\
	\]
	{\rm which implies that}
	\begin{align} \label{eq:DefEqu}
		\rres_{x_2}(\Eval{x_1,\sigma_1}(f),\Eval{x_1,\sigma_1}(g)) &= \Det(x_2^{m-1}
		\Eval{x_1,\sigma_1}(f), \ldots, x_2 \Eval{x_1,\sigma_1}(f), \Eval{x_1,\sigma_1}(f),
		\nonumber \\
		& ~~~~\, \,\, x_2^{n-1} \Eval{x_1,\sigma_1}(g),
		 \ldots, x_2 \Eval{x_1,\sigma_1}(g),
		 \, \Eval{x_1,\sigma_1}(g) ).
	\end{align}
\end{defn}

\noindent
Similar to its commutative counterpart~\cite{Collins1971}, we can prove
the following theorem.

\begin{thm} \label{thm:EvalRes=ResEval}
	Let $\Sk=\tilde{\A}[x_1;\sigma_1][x_2;\sigma_2]$ be an Ore algebra.
	For all polynomials $f,g \in \Sk$, if
	$\deg_{x_2}(f)=\deg_{x_2}(\Eval{x_1,\sigma_1}(f))$
	and $\deg_{x_2}(g)=\deg_{x_2}(\Eval{x_1,\sigma_1}(g))$,
	where $\deg_{x_2}$
	is the largest power of $x_2$ whose coefficient is not zero,
	then the following formula holds:
	\begin{equation} \label{eq:missing_link}
		\Eval{x_1,\sigma_1}(\rres_{x_2}(f,g)) =
		\rres_{x_2}(\Eval{x_1,\sigma_1}(f),\Eval{x_1,\sigma_1}(g)).
	\end{equation}
\end{thm}

\pf Let $f, g \in \Sk$ be two bivariate Ore polynomials of positive
degrees $n$ and $m$ respectively. By its definition, we can write the
right side of Equation(\ref{eq:missing_link}) as:
\[
	\begin{aligned}
		\rres_{x_2}(\Eval{x_1,\sigma_1}(f),\Eval{x_1,\sigma_1}(g)) &= \Det(
		~x_2^{m-1} \Eval{x_1,\sigma_1}(f), \ldots, x_2 \Eval{x_1,\sigma_1}(f),
		\, \Eval{x_1,\sigma_1}(f), \,\, ~\\
		 &\quad x_2^{n-1} \Eval{x_1,\sigma_1}(g), \ldots, x_2 \Eval{x_1,\sigma_1}(g),
		 \, \Eval{x_1,\sigma_1}(g)) ~\\[0.5em]
		&= \Det(
		\Eval{x_1,\sigma_1}(x_2^{m-1} f), \ldots, \Eval{x_1,\sigma_1}(x_2
		f), \, \Eval{x_1,\sigma_1}(f), \,\, ~\\
		  &\quad ~\Eval{x_1,\sigma_1}(x_2^{n-1} g), \ldots, \Eval{x_1,\sigma_1}(x_2 g),
		 \, \Eval{x_1,\sigma_1}(g)) ~\\[0.5em]
	&= \Eval{x_1,\sigma_1}( \, \Det(
	x_2^{m-1} f, \ldots, x_2 f, \, f, \,\,
		x_2^{n-1} g, \ldots, x_2 g, \, g )) ~\\[0.5em]
		&= \Eval{x_1,\sigma_1}(\rres_{x_2}(f,g)).
		 ~\\
		\end{aligned}
\]

\noindent
Note that in the last step we have assumed that the degrees of the polynomials
stay the same since it is given that the degrees will be preserved after
evaluation. This completes the proof.

\EOP ~\\

Using Theorem~\ref{thm:EvalRes=ResEval}, we can conclude that the two
methods $\Eval{x_1,\sigma_1}(\rres_{x_2}(f,g))$ and
$\rres_{x_2}(\Eval{x_1,\sigma_1}(f),\Eval{x_1,\sigma_1}(g))$ are the
same (viewed as operators). Thus, for all $a$ in $\A$ we have:
\begin{equation} \label{eq:Evala}
	\Eval{x_1,\sigma_1}(\rres_{x_2}(f,g))(a) =
	\rres_{x_2}(\Eval{x_1,\sigma_1}(f),\Eval{x_1,\sigma_1}(g))(a).
\end{equation}

The left side of Equation~(\ref{eq:Evala}) describes the operator
evaluation of direct resultant of two bivariate Ore polynomials $f$
and $g$ at a value $a$ in $\A$, while the right side provides a way
how to obtain the resultant via operator evaluation of its entries
which follows by applying evaluation at $a$.

\begin{remk}
	An advantage of using Dieudonn{\'e} determinant in
	Theorem~\ref{thm:EvalRes=ResEval} is that the case can be reduced to
	a triangular determinant with diagonal entries of polynomials
	$d'_i(x_1)$ \textup($i=1,\ldots, k; \, k=n+m$\textup) for the direct
	method of the left side of Equation~(\ref{eq:Evala}), while the right
	side will be in the form
	$d_i(\sigma_1)$ \textup($i=1,\ldots, k; \, k=n+m$\textup)
	which can be computed by the following product:
	\begin{align} \label{eq:comp}
		\rres_{x_2}(\Eval{x_1,\sigma_1}(f),\Eval{x_1,\sigma_1}(g))(a)
		&= (\prod_{i=1}^{k} d_i(\sigma_1))(a) \nonumber \\
		&= (d_1(\sigma_1)d_2(\sigma_1) \cdots d_k(\sigma_1))(a) \nonumber \\
		&= d_1^{*}(d_2^{*}( \cdots d_k^{*}(a))),
	\end{align}
	where $d_i^{*} = d_i(\sigma_1)$ for all $i=1, \ldots, k $.
\end{remk}

It is evident that over some finite division rings (finite fields),
computing the product~(\ref{eq:comp}) by evaluation and interpolation
methods~\cite{Eric2020,Caruso2017} is asymptotically more efficient
than its corresponding product by the direct method. Thus, we can
consider the evaluation and interpolation method for the bivariate Ore
polynomial rings over finite fields.

The idea of the evaluation and interpolation method is that instead of
computing resultant directly from two bivariate Ore polynomials
$f_1(x_1,x_2)$ and $f_2(x_1,x_2)$ with respect to $x_2$, we propose to
chose enough distinct values $a_i \, (i=0,\ldots,k)$ to compute the
evaluation of $f_1$ and $f_2$ with respect to $x_1$ at the values
$a_i$ $(i=0,\ldots,k)$. Then, we will compute what we call
{\em partial resultants} of these already evaluated $f_1$ and $f_2$
with respect to $x_2$. Finally, we can deduce the original resultant by
combining all these partial resultants using an interpolation technique.
However, this process is not straightforward in Ore algebra, some
challenges need to be overcome:

\begin{enumerate}[align=left, label=(\roman*)]
    \item The current process of the right side of Equation~(\ref{eq:Evala}) is to evaluate
        a polynomial at $\sigma_1$ and then applying the whole
        operator polynomial to a selected value of $a$ in $\A$, which may not reflect the
        true evaluation of the original resultant at the left side of the
        equation unless applied in the same exact manner, and this makes
        it difficult to recover the original polynomial through a Lagrange like
        interpolation technique without modifications or
        by using another evaluation and interpolation method.
        One way to overcome this difficulty is using the coefficient comparison
        method which will always work in this case.
    \item In some cases, we may not even have enough values to check for
        valid evaluation points, which make it not possible to continue without
        extending the available choices. Thus, we extend the domain
        to enough elements for the interpolation process. Recall, we have
        the freedom to choose
        these evaluation points (as long as they belong to distinct conjugacy classes).
    \item The assumption of preserving the degrees of the input polynomials
        before and after evaluation in Theorem~\ref{thm:EvalRes=ResEval},
        is to avoid what is called {\em bad evaluation} where an
        evaluation value could cause the leading coefficient to vanish,
        hence altering the degree of the polynomial.
        Therefore, we need to check whether an evaluation value is a bad
        evaluation which can easily be determined from the leading term.
    \item We need to make sure all the selected evaluation values belong
        to pairwise distinct conjugacy classes in order to be able to
        use evaluation and interpolation techniques. For this, we can
        choose primitive elements and show that they belong to different
        conjugacy classes.
\end{enumerate}

\section{Conclusion} \label{sec:conclusion}

In this work, we used an elimination technique for bivariate Ore polynomials
by using new resultant computations directly from the polynomial coefficients,
as well as by using evaluation and interpolation methods. The methodology
uses modular approaches to further optimize the algorithms.

Also, to conclude that the challenges mentioned in the introduction section
can be overcome as following:

\begin{enumerate}[label=(\roman*)]
	\item \label{item:Challenges-iv}  {\em No resultant for bivariate
		Ore polynomials}: We have defined a new resultant
		for bivariate Ore polynomials (including the case where the entries
		are automorphisms).
	\item \label{item:Challenges-i} {\em Evaluation map is not common}:
		In general, it is known that most of the noncommutative evaluation
		maps are not ring homomorphism with the exception
		of operator evaluation~\cite{Ulmer2013} in which we had to extend it to the
		bivariate case.
	\item \label{item:Challenges-iii}  {\em Noncommutative determinant
		is different}: We have used Dieudonn{\'e} determinant to compute
		the resultant which is unique modulo commutators.
	\item \label{item:Challenges-ii} {\em Interpolation requires more
		conditions}: The selected values for the noncommutative interpolation
		need to be in distinct conjugacy classes. We can extend the domain
		to distinct primitive elements which then can be used for the
		the evaluation and interpolation method.
\end{enumerate}

\section*{Acknowledgment} \label{sec:acknowledgment}

The author wishes to thank Dr.~Yang~Zhang for introducing him to
Ore polynomials and for comments on some parts of this paper, and
Dr.~Parimala~Thulasiraman for comments on earlier versions.~The author
bears the sole responsibility for the content of this study.

\end{document}